\documentclass{JINST}

\title{CMS reconstruction improvement for the muon tracking by the RPC chambers}

\author{Min Suk Kim\thanks{Corresponding author.}~ on behalf of the CMS Collaboration\\
\llap{}Sungkyunkwan University, Department of Physics,\\
  2066, Seobu-ro, Jangan-gu, Suwon, Gyeonggi-do, Republic of Korea\\

E-mail: \email{minsuk@cern.ch}

}

\abstract{
The contribution of Resistive Plate Chamber (RPC) to muon reconstruction in CMS has been studied on a sample of muons collected in proton-proton collisions at $\sqrt s$ = 7 TeV at the LHC in 2011.
Muon reconstruction is performed using the all-silicon inner tracker and with up to four stations of gas-ionization muon detectors. 
Drift Tubes and Cathode Strip Chambers detect muons in the barrel and endcap regions, respectively, and are complemented by the RPC system. 
Measured distributions of reconstructed hits in the RPCs crossed by muons from $Z^{0}$ decays with a transverse momentum $p_T$ above 20 GeV/\emph{c} are well reproduced by the Monte Carlo simulation.
From the samples of $J/\psi$ and $Z^{0}$ events, the efficiencies for muons with and without the inclusion of the RPC hits in the muon track reconstruction are measured and compared with the simulation.
Using RPC information in track reconstruction improves up to about 3$\%$ of offline reconstruction efficiency for the muons in the region of $p_T$ above 7 GeV/\emph{c}, in good agreement with simulation.
}


\begin{document}

\section{Introduction}

The key to new discoveries and precision measurements at the Large Hadron Collider (LHC) is the ability of excellent muon identification, triggering and precise momentum reconstruction with high efficiency.
The muon system of the Compact Muon Solenoid (CMS) experiment~\cite{CMS_experiment} and the muon reconstruction algorithms have been designed to achieve these goals.
The focus of this paper is the contribution of the RPC sub-system to the muon identification and reconstruction capabilities, alongside the Drift Tubes (DT) and Cathode Strip Chambers (CSC) independent sub-systems.

Muon reconstruction is performed using the central tracker and with all three muon detector systems.
Due to the redundancy provided by the combination of DT (CSC) and RPC in the barrel (endcap) region, 
it is expected that the inclusion of the RPC measurements in the muon reconstruction 
can improve the reconstruction of low momentum muons, especially in the case where the muons escape through the gaps between the wheels in which the chambers are divided along the axis parallel to the beam,
leaving hits in only one DT station.
Thus, the analysis presented here shows the impact of the RPC hits on the offline reconstruction
of the muons from the resonance decays of $J/\psi \rightarrow \mu^+\mu^-$ events in the region of $p_T$ below 20 GeV/\emph{c} and $Z^{0} \rightarrow \mu^+\mu^-$ events in the region of $p_T$ above. 
Furthermore, the number of reconstructed hits in the RPC chambers which could be associated to a muon track has been compared to the Monte Carlo (MC) expectations.

\section{Detector Layout}

\begin{figure}
  \begin{center}
    \texttt{\includegraphics[width=0.66\textwidth]{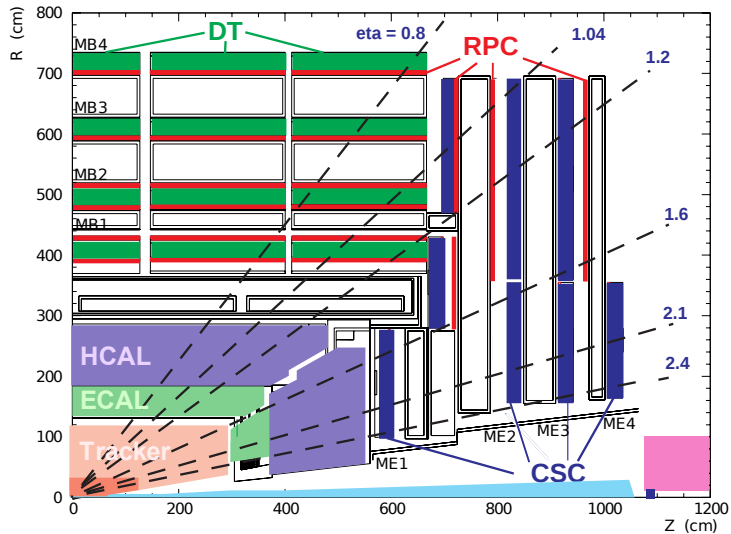}}
    \caption{
    Layout of one quadrant of CMS. The four DT stations in the barrel (MB1-MB4, green), the four CSC stations in the endcap (ME1-ME4, blue), and the RPC stations (red) are shown.
    }
    \label{fig:MuonSys}
    \vspace{-0.5cm}
  \end{center}
\end{figure}

A schematic view of the CMS detector is shown in Figure~\ref{fig:MuonSys}.
The muon system covers the pseudo-rapidity region $|\eta|$ < 2.4~\cite{MUO-10-004}. DTs and CSCs detect muons in the regions of $|\eta|$ < 1.2 and 0.9 < $|\eta|$~<~2.4, respectively, and are complemented by the RPC system covering the range~of~$|\eta| $~<~1.6~\cite{RPC_chambers}.

For the study of the RPC contribution to muon reconstruction, we define three~regions in the detector, referred to as barrel ($|\eta|<0.8$), transition ($0.8<|\eta|<1.2$), and endcap ($1.2<|\eta|<1.6$).
The current design of the RPC endcap is not expected to be suitable for high particle rates at high~$\eta$ in the scenario of an LHC luminosity going up to $10^{34-35}~{\mathrm{cm}}^{-2} \cdot {\mathrm{s}}^{-1}$. There is ongoing work in identifying suitable technologies to instrument the high $\eta$ region 
up to 2.4 matching the CSC system.
The RPCs are used at trigger level and also in the standard offline muon reconstruction.

In the barrel region the muon chambers are organised in four coaxial stations,
interleaved with iron return yokes.
A station is an assembly of chambers around a fixed value of $r$ in the barrel (or $z$ in the endcap).
Each of the two inner stations (MB1, MB2) contains layers of DTs sandwiched between two layers of RPCs, whereas each of the two outer stations (MB3, MB4) consists of one layer of RPCs and layers of DTs.
Along $z$, these barrel stations are grouped into five wheels, which are in turn divided into twelve $\phi$ sectors.
Similarly in the $r$ direction in the endcaps there are rings of CSCs, increasing with the radial distance from the beam line.
Each endcap ring is composed of 36 chambers covering the full azimuthal range.
The endcap region is composed of three iron disks holding a total of three RPC planes and four CSC planes on each side (plus/minus) along $z$. 

\section{RPC Hits in Muon Reconstruction}
Muon tracking in CMS can be performed with the silicon tracker, and with either three or four stations of muon chambers installed outside the solenoid, sandwiched between steel layers serving both as hadron absorbers and as a return yoke for the magnetic field~\cite{PLB_chargeRatio}.

Three types of muon-track reconstruction were designed for muons originating from the LHC proton-proton collisions.
Muon tracks can be reconstructed by using hits in the muon detectors alone (standalone muons). Alternatively, the reconstruction can combine hits in the muons detectors with those in the central tracker (global muons). The muon system can also be used purely to tag extrapolated tracks from the central tracker; such tracks qualify as tracker muons if at least one muon segment (i.e. a short track stub made of DT or CSC hits) matches the extrapolated track.

\begin{figure}
  \begin{center}
    \texttt{\includegraphics[width=0.8\textwidth]{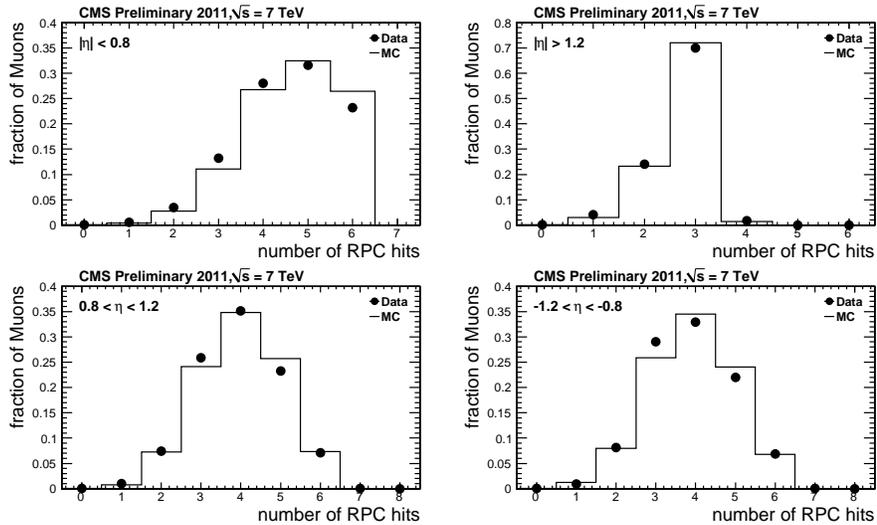}}
    \caption{
    Number of RPC hits per global muon in $Z^{0}$ events for the barrel, endcap, transition minus and plus regions (clock-wise, starting from top-left). Results are shown for both data (dots) and MC (solid line). The error bars indicate the statistical uncertainty and are smaller than the points.
    }
    \label{fig:hitDistr_STA}
    \vspace{-0.4cm}
  \end{center}
\end{figure}

\begin{figure}
  \begin{center}
    \texttt{\includegraphics[width=0.44\textwidth]{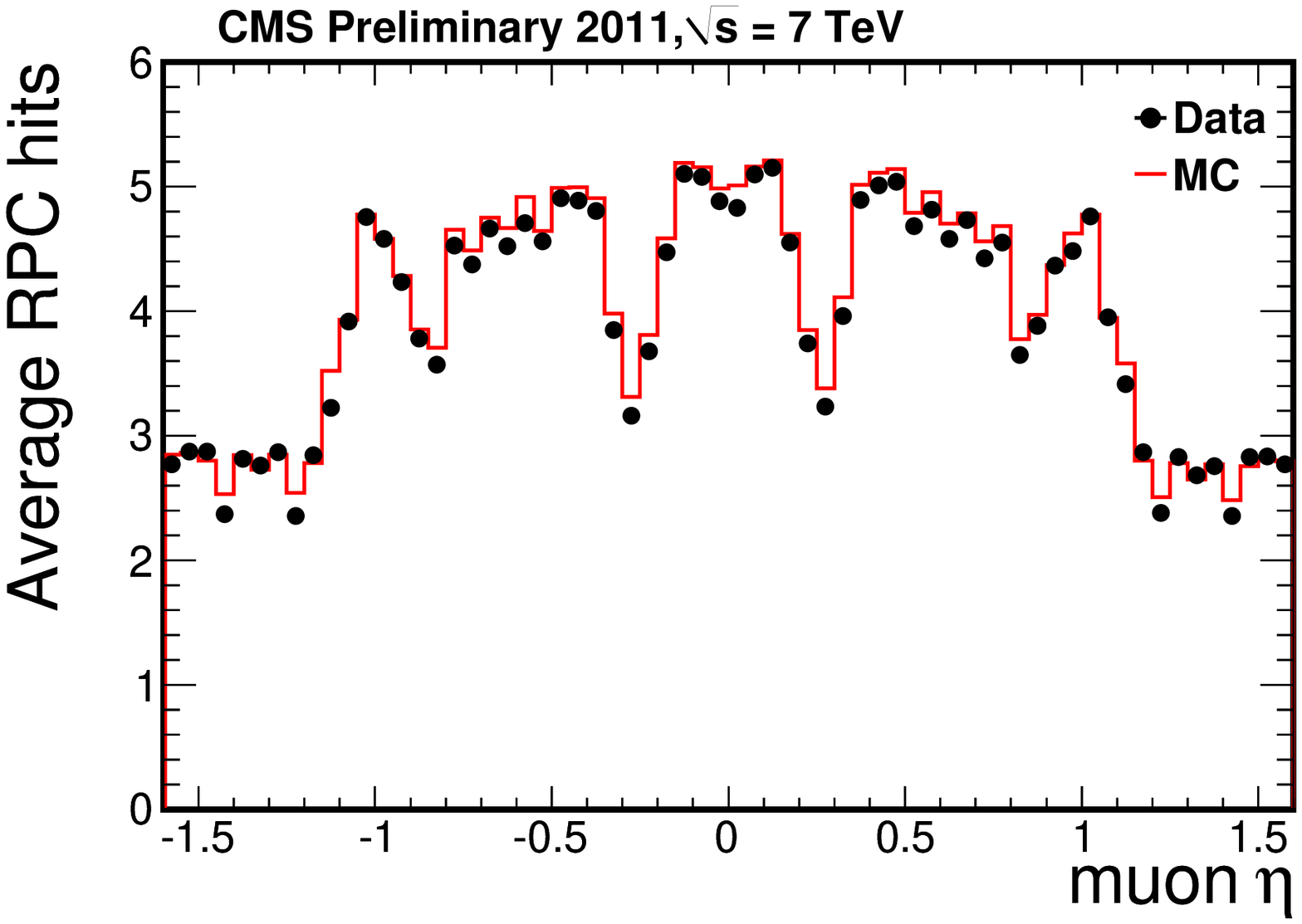}}\hspace{0.15cm}\texttt{\includegraphics[width=0.55\textwidth]{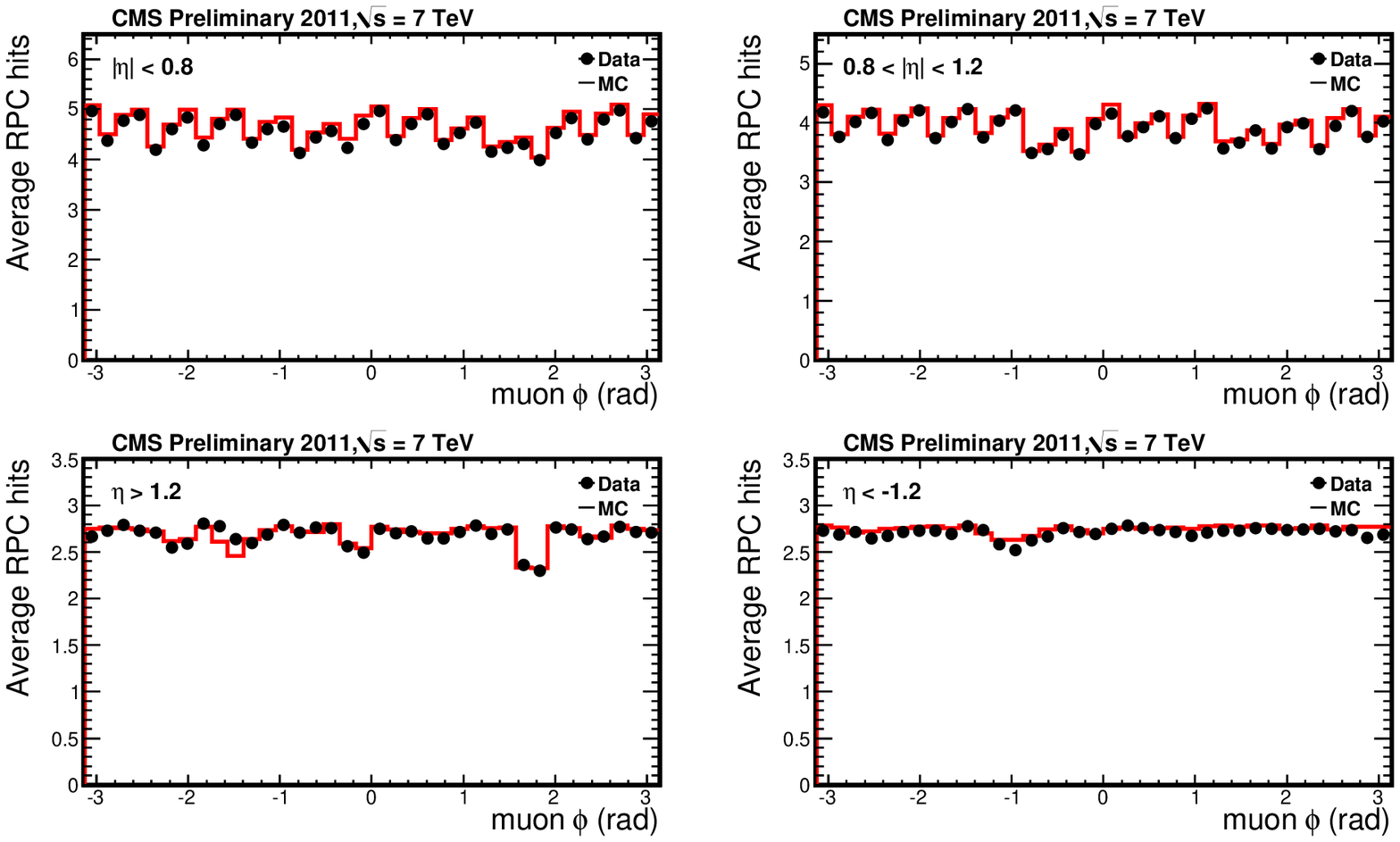}}
    \caption{
    Average number of RPC hits of global muons in $Z^{0}$ events vs. $\eta$ (left) and $\phi$ for the barrel, transition, and endcap minus and plus regions (right four plots, clock-wise, starting from top-left).  The error bars indicate the statistical uncertainty and are smaller than the points.
    }
    \label{fig:fraction_STA}
  \end{center}
\end{figure}

For the reconstruction of the standalone- or global-muon track, at least two measurements, one of which must be of the DT or CSC segment, must be matched within the muon system.
The RPC hits are used by default in the standard muon reconstruction.
Figure~\ref{fig:hitDistr_STA} shows the number of RPC hits associated to global muons for the barrel, transition and endcap regions. 
Figure~\ref{fig:fraction_STA} shows the average number of RPC hits of global muons as a function of $\eta$ and $\phi$ for given $\eta$ regions.
The dips in the barrel $\eta$ around $\pm$0.25 and $\pm$0.8 are due to the gaps between the 5 wheels of the yoke.
Similarly, the oscillations in the $\phi$ distribution for RPC hits with $|\eta|$ < 1.2 are due to the cracks between the adjacent sectors.
There is a small bias from the $Z^{0}$ event selection with at least one muon triggered.
The error bars indicate the statistical uncertainty and are smaller than the points.

\section{Muon Efficiencies including or excluding RPC hits}
In this study we measure the efficiency of the selections
with given $p_T$ thresholds for low (3.5 < $p_T$ < 7 GeV/\emph{c} and 7 < $p_T$ < 20 GeV/\emph{c})
and high $p_T$ muons ($p_T$ > 20 GeV/\emph{c}) based on the following muon identification variables. $\chi^{2}$ divided by degrees of freedom in the global-muon track fit is less than 10, and its corresponding tracker track must use more than 10 silicon tracker hits
and have a transverse impact parameter $|d_{xy}| <$ 2 mm with respect to the primary vertex.
The minimal number of measurement points in the tracker is also required for the muons from the $J/\psi$ decays.

\begin{figure}
  \begin{center}
    \texttt{\includegraphics[width=0.45\textwidth]{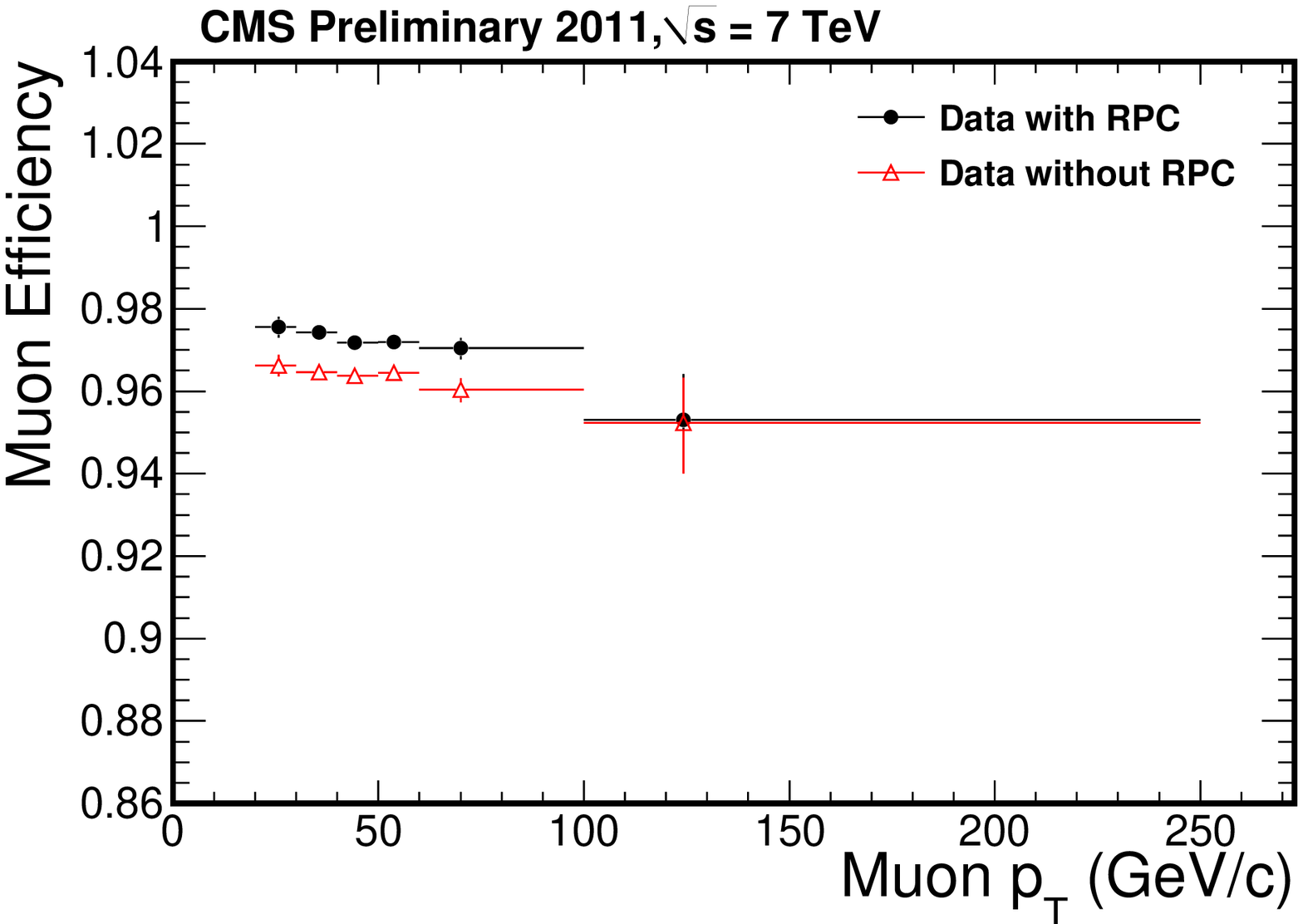}}\hspace{0.5cm}\texttt{\includegraphics[width=0.45\textwidth]{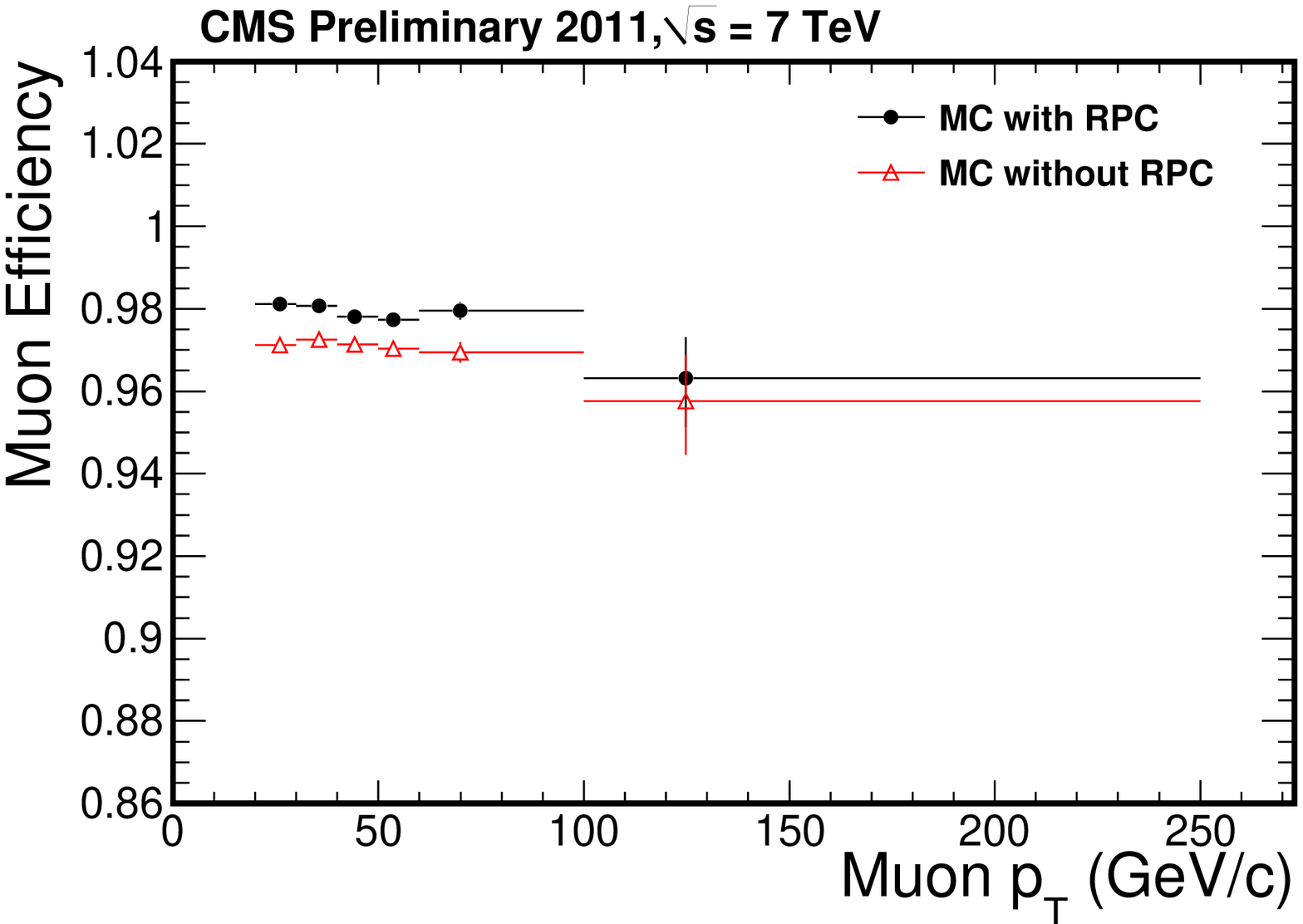}}
    \texttt{\includegraphics[width=0.45\textwidth]{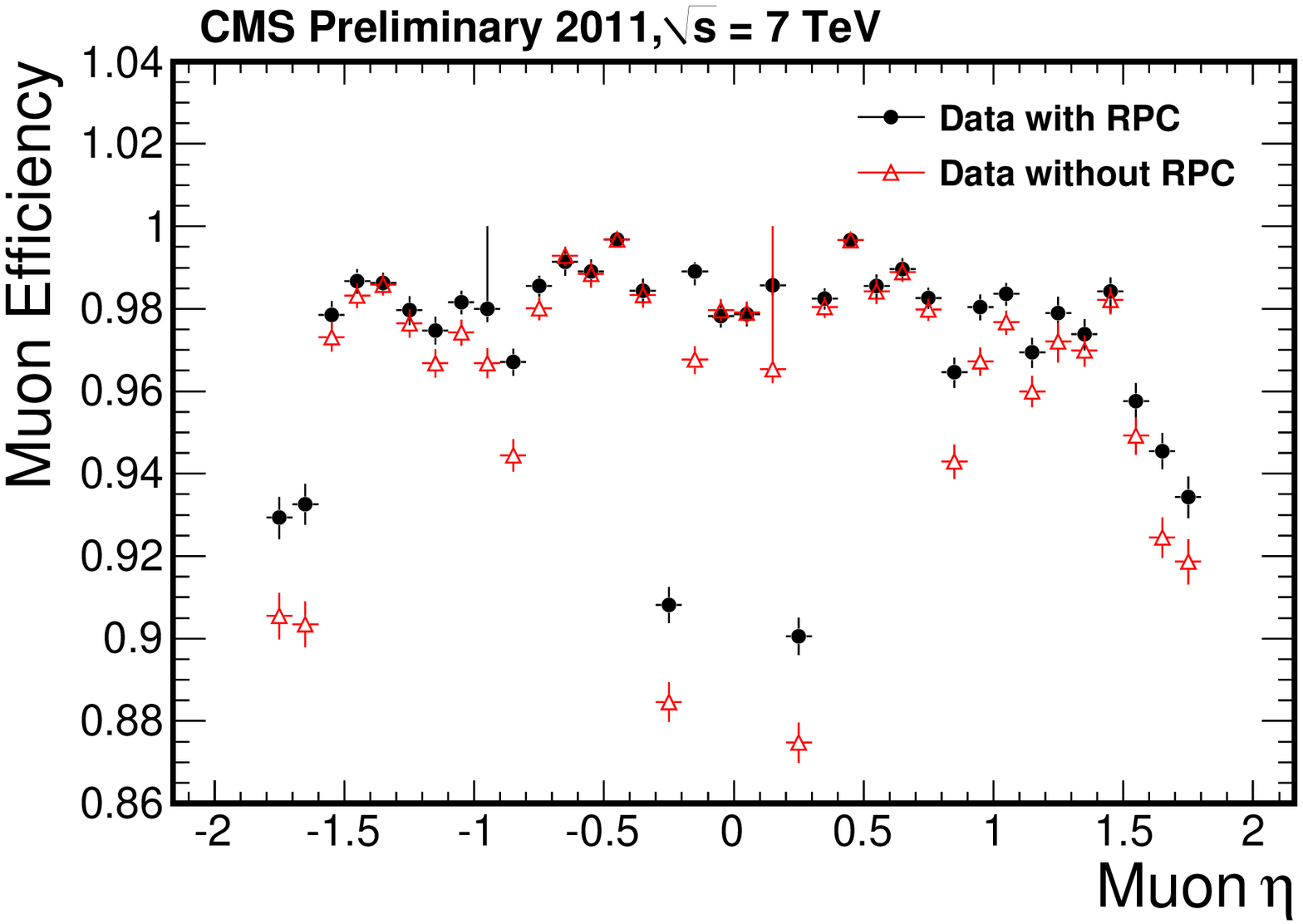}}\hspace{0.5cm}\texttt{\includegraphics[width=0.45\textwidth]{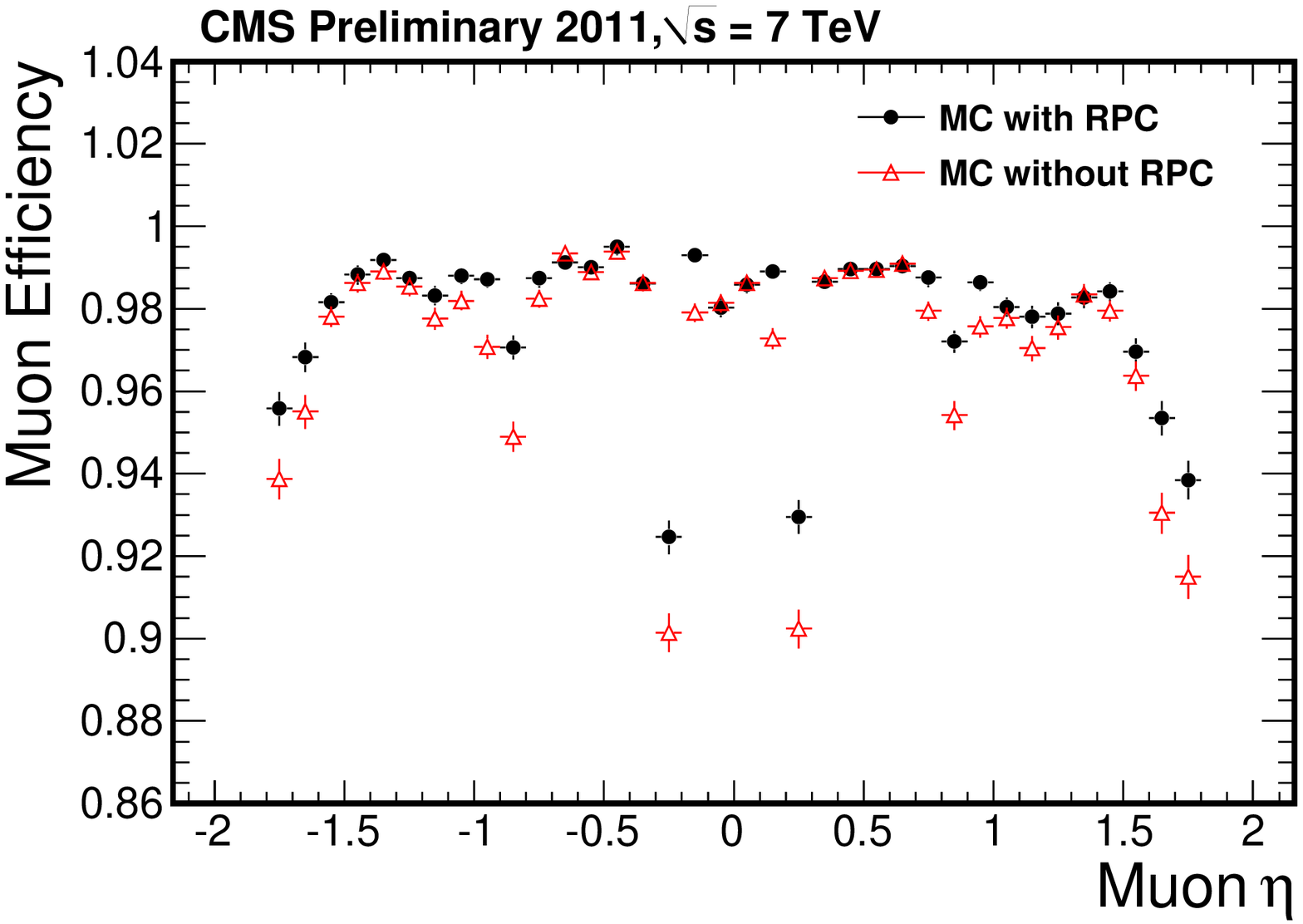}}
    \caption{
    For $Z^{0}$ events, efficiencies measured on data (left) and MC (right) as a function of $p_T$ (top) and $\eta$ (bottom).
    The black points are the global muon in the standard reconstruction ($\epsilon_{\mathrm{withRPC}}$) and the red open triangles are the global muon reconstruction by using hits in the tracker and DT/CSC only ($\epsilon_{\mathrm{withoutRPC}}$).
    }
    \label{fig:muEffAbs}
    \vspace{-0.4cm}
  \end{center}
\end{figure}

A global-muon track can be fitted combining hits from the tracker track and standalone-muon track, using the Kalman-filter technique~\cite{Kalman}.
The efficiency for global muons in the standard CMS reconstruction that uses RPC hits in the global fit ($\epsilon_{\mathrm{withRPC}}$) was measured and compared with the efficiency for muon reconstruction when RPC hits are excluded from the global fit ($\epsilon_{\mathrm{withoutRPC}}$).
In this way the RPC contribution to the global muon reconstruction efficiency is evidenced.


For these efficiency measurements, one can use the events with well known resonance decays into a pair of muons
and consider one of the muons as the tag to probe the efficiency using the other muon.
The tag is a well isolated muon, while the probe has less stringent selection requirements without isolation. We used the samples of $Z^{0}$ and $J/\psi$ events corresponding to an integrated luminosity of about $1~\rm{fb^{-1}}$ and $4~\rm{fb^{-1}}$, respectively: the simulated processes are $Z^{0} \rightarrow \mu^+\mu^-$ and $J/\psi \rightarrow \mu^+\mu^-$ only.
Events are selected from samples collected using the high-$p_T$ single-muon trigger 
and the dedicated $J/\psi$ triggers to enrich the purity of the low-$p_T$ single-muon trigger by selecting events in which the muon track can be paired to a tracker track of opposite charge yielding an invariant mass close to that of the $J/\psi$ peak.
The tag is a global muon with the following additional requirements: associated with at least two track segments in different muon stations and at least one muon chamber hit included in the global-muon track fit. To further purify the sample the isolation cuts based on the tracker and calorimeters have been applied only for high-$p_T$ muons.

To measure efficiency for high-$p_T$ muons, the probe is a tracker track fulfilling the following selection criteria,
$p_{T}$ > 20 GeV/\emph{c}, $|\eta| <$ 1.8 and at least 10 hits in the tracker, so there is no bias from the muon sub-detectors.
The tag and probe tracks are also required to share the same primary vertex.
The invariant mass of the tag and probe tracks is fitted to extract the signal event yields for both the denominator and the numerator on the efficiency ratio, requiring all probes in the numerator match a global muon candidate reconstructed with and without 
the RPC, respectively. 

\begin{figure}
  \begin{center}
    \texttt{\includegraphics[width=0.45\textwidth]{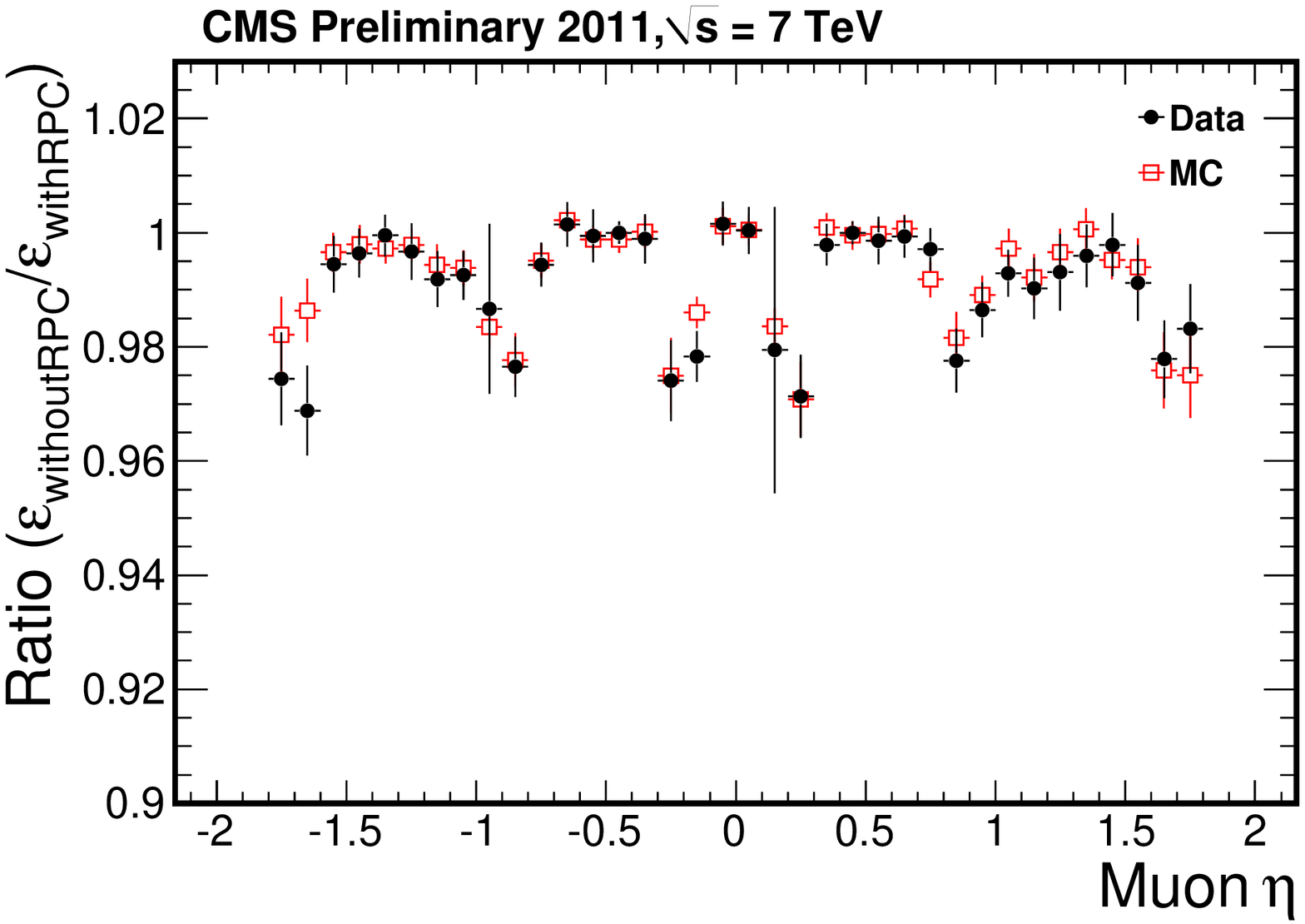}}\hspace{0.5cm}\texttt{\includegraphics[width=0.45\textwidth]{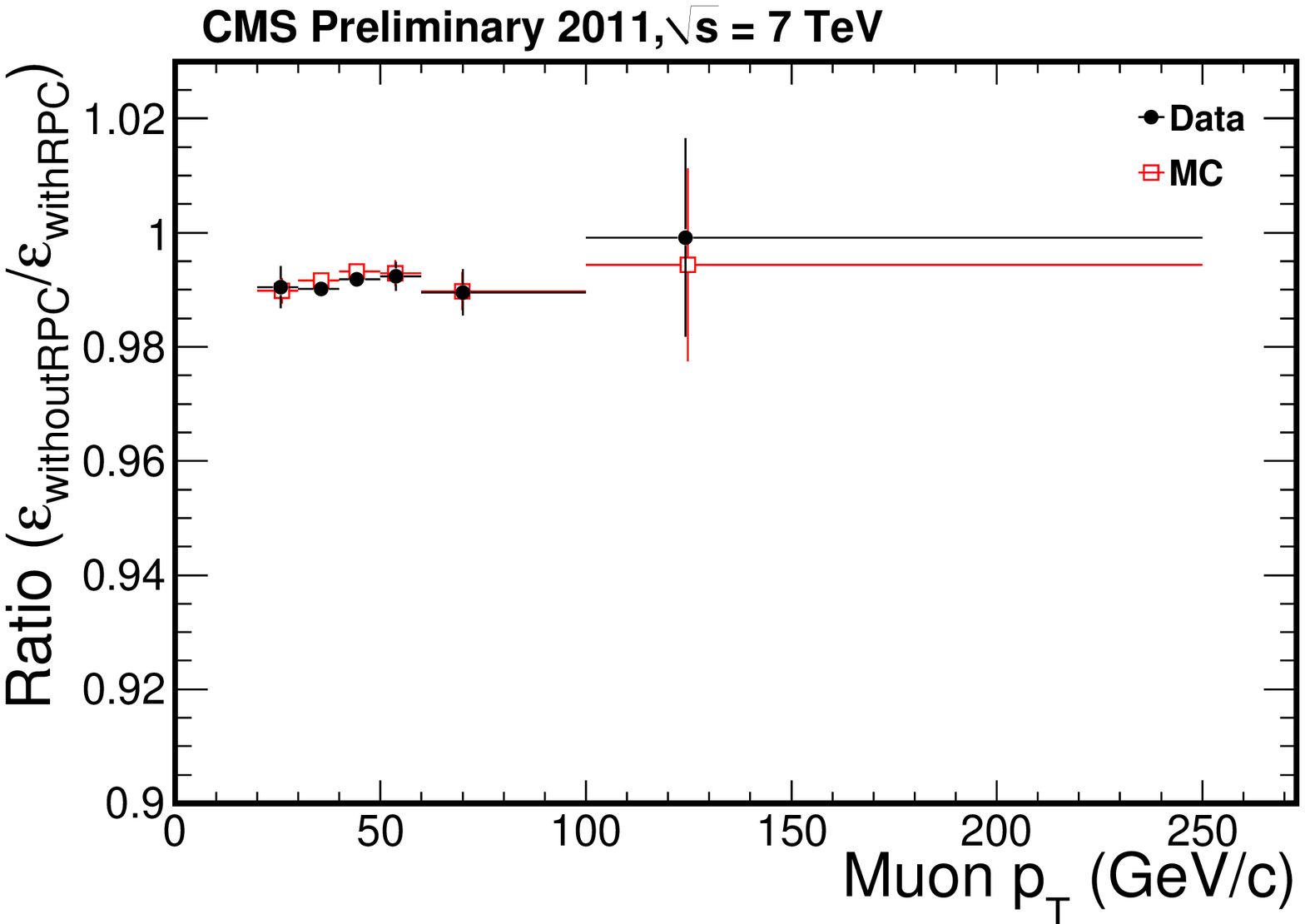}}
    \caption{
    For $Z^{0}$ events, the ratio between the efficiencies with and without the inclusion of the RPC hits in the global muon reconstruction ($\epsilon_{\mathrm{withoutRPC}}/\epsilon_{\mathrm{withRPC}}$).
Results are shown for both data (dots) and MC (red open squares) as a function of $\eta$ (left) and $p_T$ (right), respectively.
    }
    \label{fig:muEffRatio}
    \vspace{-0.3cm}
  \end{center}
\end{figure}

Figure~\ref{fig:muEffAbs} shows
the efficiency of the muons coming from $Z^{0}$ decays, for both data and MC samples,
by applying a tag-and-probe technique to obtain almost unbiased estimates.
The standard global muon reconstruction with RPC is more efficient than the global muon reconstruction without RPC over the full $p_T$ and $\eta$ ranges, because it requires at least one muon segment if a segment matches at least one RPC hit within the muon system.
The efficiency drops with excluding the RPC in the global muon reconstruction, 
which varies from a few percent to 3\% depending on $\eta$. 

The ratio between the efficiencies for global muons reconstructed with and without the RPC ($\epsilon_{\mathrm{withoutRPC}}/\epsilon_{\mathrm{withRPC}}$) is shown in Figure~\ref{fig:muEffRatio}
as a function of $\eta$ and $p_T$, respectively.
The results on data are compared with the ones extracted applying the same procedure on MC samples.
Overall, there is good agreement between data and simulation in both the efficiencies and shapes of the distributions. Some discrepancies result from imperfect simulation of local detector conditions, affecting for example the muon identification efficiency~\cite{MUO-10-004}.

\begin{figure}
  \begin{center}
    \texttt{\includegraphics[width=0.45\textwidth]{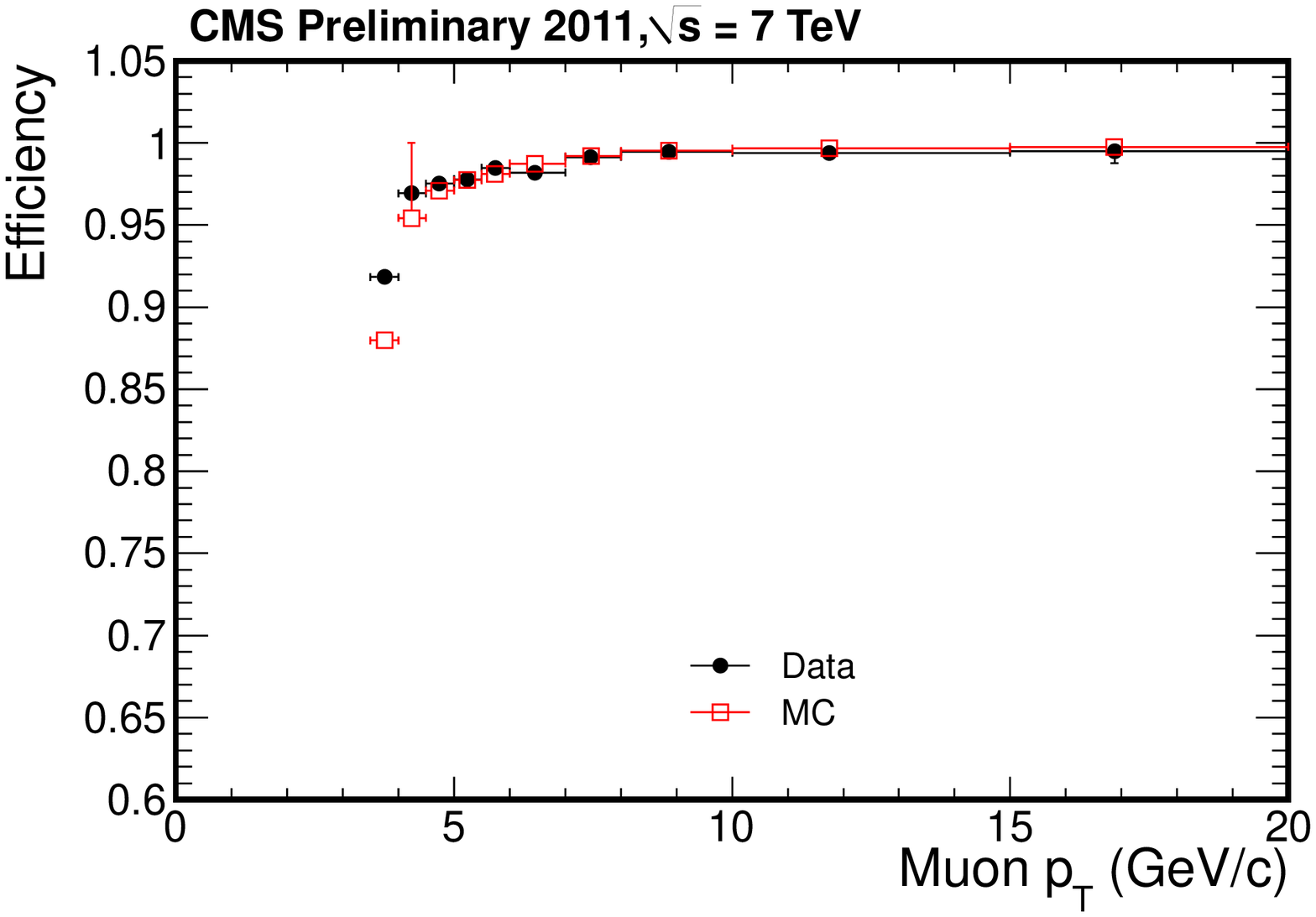}}\\
    \vspace{0.3cm}
    \texttt{\includegraphics[width=0.45\textwidth]{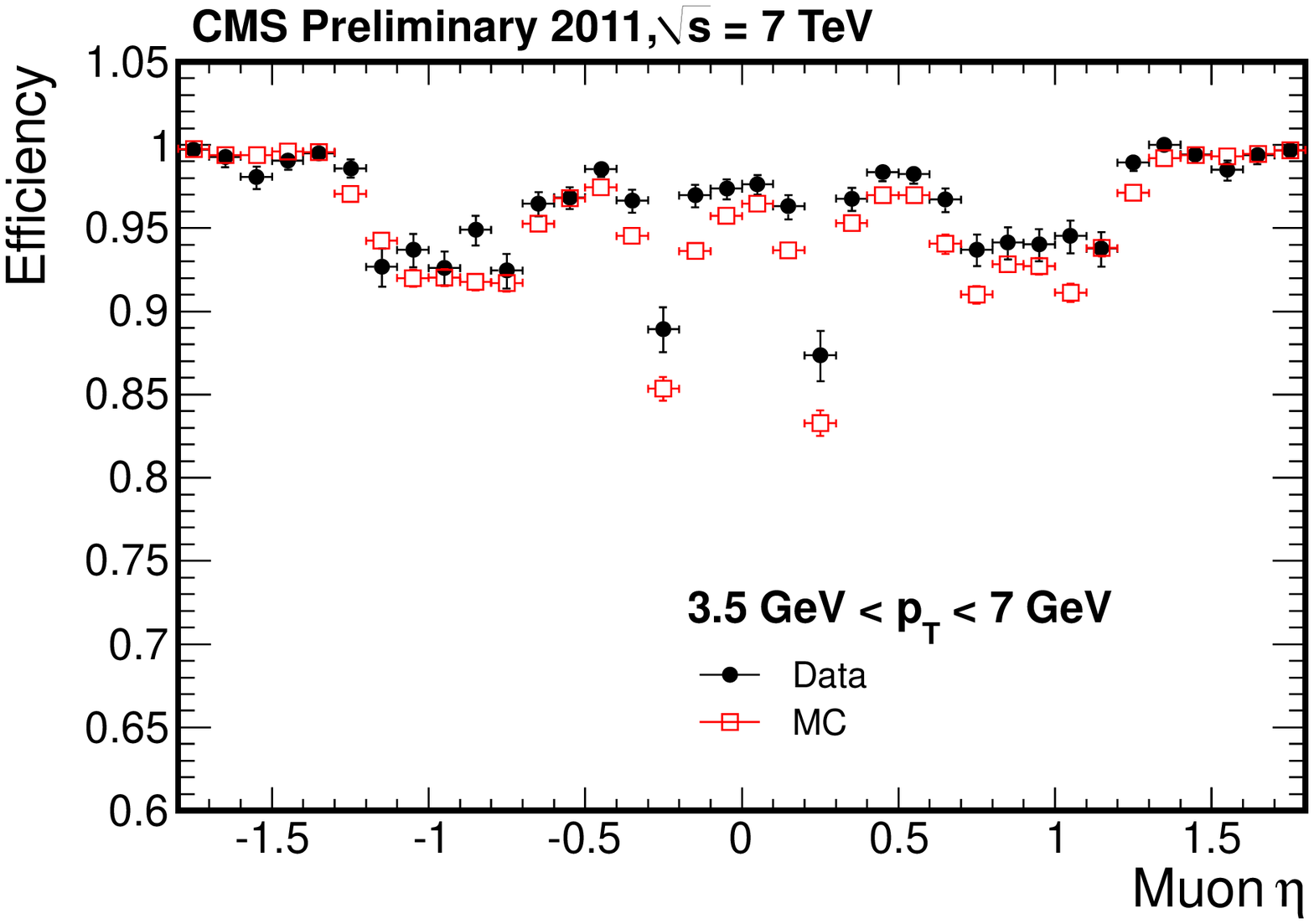}}\hspace{0.5cm}\texttt{\includegraphics[width=0.45\textwidth]{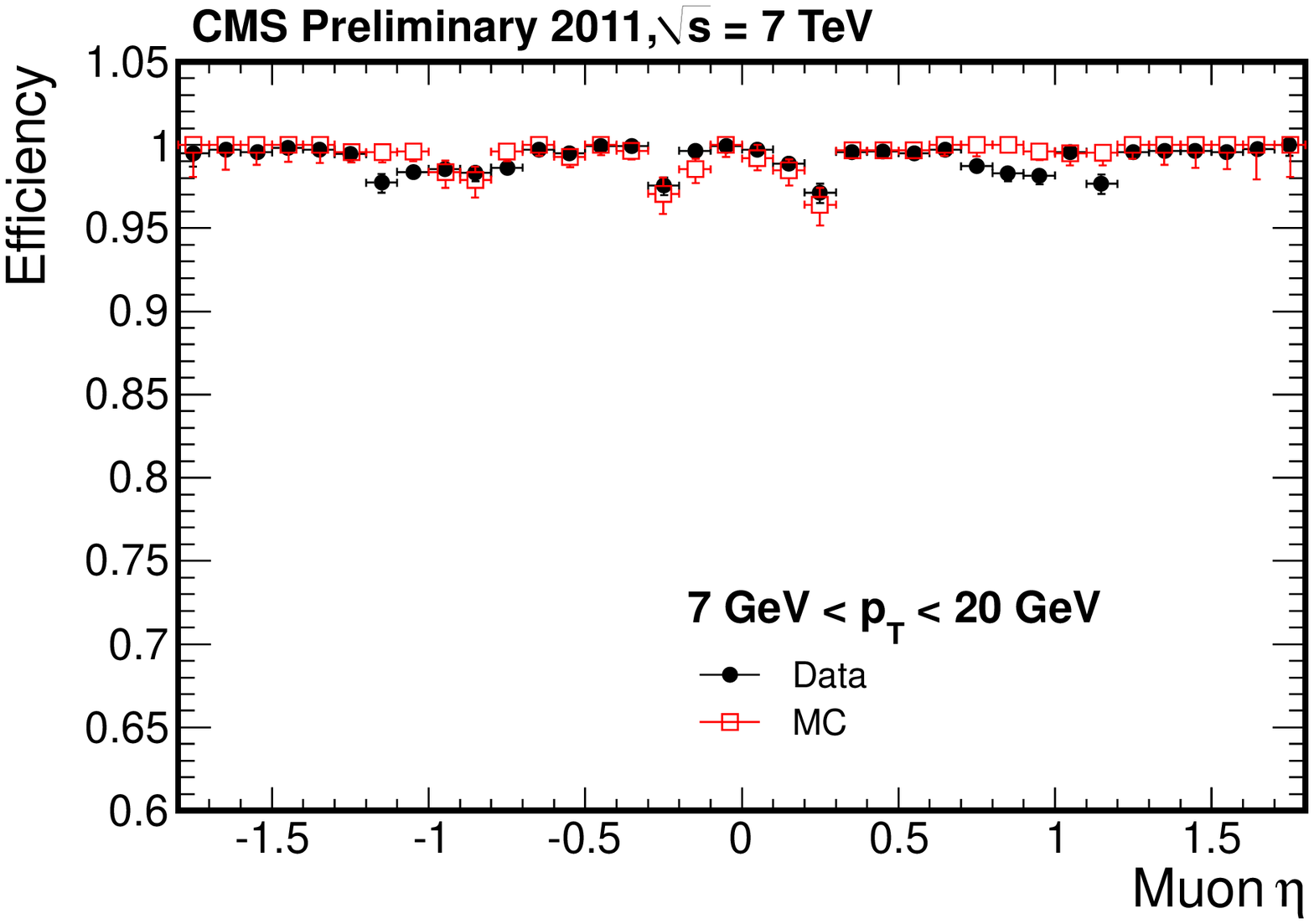}}
    \caption{
    For $J/\psi$ events, efficiency for muons that have been reconstructed with the standard algorithm (including RPC hits) to also be reconstructed once RPC hits are excluded.
Results are shown as a function of $p_T$ (top) and $\eta$ (bottom) for both data (dots) and MC (red open squares). The bottom plots show efficiencies for muons with 3.5 < $p_T$ < 7 GeV/\emph{c} (left) and 7 < $p_T$ < 20~GeV/\emph{c} (right), respectively, in the events selected by $J/\psi$ triggers with different thresholds on $p_T$.
    }
    \label{fig:muEffRel}
    \vspace{-0.3cm}
  \end{center}
\end{figure}

In the case of the $J/\psi$ events, combinatorial backgrounds from other tracks in the event are high.
Hence, for the muon efficiency at low $p_T$, the probe is a global muon with at least 10 hits in the tracker.
Figure~\ref{fig:muEffRel} shows the efficiency for muons that have been reconstructed with the standard algorithm (including RPC hits) to also be reconstructed once RPC hits are excluded.
The top plot shows the $p_T$ dependence of efficiency and the only significant discrepancy is around the turn-on of the efficiency curves, where the efficiency in data is systematically higher than that in the simulation (in the region $|\eta|$ < 1.2 as shown in the bottom-left plot). This discrepancy arises from a difference in the widths of the track-to-segment pulls between data and simulation.
The bottom plots show efficiencies for muons 
in the events selected by $J/\psi$ triggers with different thresholds~on~$p_T$.

Some of tag-probe pairs may come from the resonance decays, while  other tag-probe pairs will be combinatorial background, where the probe is usually a charged hadron.
This combinatorial background can be removed in a simultaneous fit to the invariant mass spectra for passing and failing probes with identical signal shape and appropriate background shapes. The normalizations of the signal shapes in the two spectra are used to compute the efficiency.

The uncertainty on the fitted efficiency is determined from the likelihood function.
As normalizations of signal and background, efficiency of the background, and parameters controlling the shapes of the signal and background are all parameters of the fit, the uncertainty includes the contributions from the background subtraction procedure.
When the background is large, as in the lower mass resonances,
the uncertainty on the efficiency obtained by the fit is purely statistical~\cite{MUO-10-004}.
The typical statistical uncertainty on the $J/\psi$ results presented here
was estimated to be about $1\%$. 

As a result, exclusion of the RPC hits in the global muon reconstruction degrades the muon efficiency by 1$\%$ on average, and to about 3$\%$ in the gaps between the wheels corresponding to certain eta region,
with tracker tracks from $Z^{0}$ decays for $p_T$ > 20 GeV/\emph{c} identified as possible muon candidates by applying the tag-and-probe method.
We see an overall good agreement between data and MC simulation within the
uncertainties.
Similar results are obtained with the muons coming from $J/\psi$ decays with 7 < $p_T$ < 20 GeV/\emph{c}.
Preliminary results on collision data at higher luminosities were checked and differences were found to be negligible.

\section{Conclusions}

The standard CMS muon reconstruction is performed using the central tracker and with all three muon detector systems: DT, CSC, and RPC.
The RPC contribution to muon reconstruction has been studied using
data collected in proton-proton collisions at $\sqrt s$ = 7 TeV during the 2011 LHC run. Tag-and-probe techniques were used to measure offline reconstruction efficiency for the muons in the region of $p_T$ above 3.5 GeV/\emph{c} using the well known resonance decays.
Using RPC hits in the muon track reconstruction improves efficiency of reconstructing $J/\psi$ and $Z^{0}$ events with muons in the gaps between the 5 wheels of the yoke (at $|\eta|$ = 0.25 and 0.8) and the transition region between the barrel outer wheels and the endcap disks ($0.8<|\eta|<1.2$).
The efficiency for reconstructing muons in the region of $p_T$ above 7 GeV/\emph{c} improves up to about 3$\%$ by the RPC chambers, in good agreement with simulation.

\acknowledgments
We congratulate our colleagues in the CERN accelerator departments for the excellent performance of the LHC machine. We thank the technical and administrative staff at CERN and other CMS institutes, and acknowledge support from BMWF and FWF (Austria); FNRS and FWO (Belgium); CNPq, CAPES, FAPERJ, and FAPESP (Brazil); MEYS (Bulgaria); CERN; CAS, MoST, and NSFC (China); COLCIENCIAS (Colombia); MSES (Croatia); RPF (Cyprus); MoER, SF0690030s09 and ERDF (Estonia); Academy of Finland, MEC, and HIP (Finland); CEA and CNRS/IN2P3 (France); BMBF, DFG, and HGF (Germany); GSRT (Greece); OTKA and NKTH (Hungary); DAE and DST (India); IPM (Iran); SFI (Ireland); INFN (Italy); NRF and WCU (Korea); LAS (Lithuania); CINVESTAV, CONACYT, SEP, and UASLP-FAI (Mexico); MSI (New Zealand); PAEC (Pakistan); MSHE and NSC (Poland); FCT (Portugal); JINR (Armenia, Belarus, Georgia, Ukraine, Uzbekistan); MON, RosAtom, RAS and RFBR (Russia); MSTD (Serbia); SEIDI and CPAN (Spain); Swiss Funding Agencies (Switzerland); NSC (Taipei); ThEP, IPST and NECTEC (Thailand); TUBITAK and TAEK (Turkey); NASU (Ukraine); STFC (United Kingdom); DOE and NSF (USA).

\end{document}